\documentclass[pra,twocolumn,amsfonts,amssymb,amsmath,floatfix,floats,a4paper]{revtex4-1}
\usepackage{float}
\usepackage[utf8]{inputenc}
\usepackage{graphicx}

\begin{document}

\title{Comment on ``Multitime quantum communication: Interesting but not counterfactual''}
\author{Lev Vaidman}
\affiliation{Raymond and Beverly Sackler School of Physics and Astronomy, Tel-Aviv University, Tel-Aviv 69978, Israel}

\begin{abstract}
In a recent paper, Robert Griffiths [Phys. Rev. A
{\bf 107}, 062219 (2023)] analyzed
a protocol for transmission of information between two parties introduced by Salih et al. [Phys. Rev. Lett. {\bf 110}, 170502 (2013)]. There is a considerable controversy about the counterfactuality of this protocol, and Griffiths suggested to resolve it by introducing a new measure of channel usage, which he called  ``Cost''. I argue that this measure is not appropriate because the original interaction-free measurement protocol which triggered the definition of the concept of counterfactuality is not counterfactual according to this measure.
\end{abstract}

%
%
%
%
%

\maketitle
Griffiths \cite{Griff} analyzed counterfactuality of the communication protocol \cite{Salih13}.
 The term `counterfactual' for describing quantum protocols was coined by Penrose \cite{Penrose} in describing interaction-free measurement (IFM) introduced by Elitzur and Vaidman \cite{IFM}:``Counterfactuals are things that might have happened, although they did not in fact happen.'' In a successful run of the IFM, the presence of an opaque object was found with the help of a probe that could have been adsorbed by the object, but actually it was not. Jozsa \cite{Jozsa}
applied this idea to `counterfactual computation', a setup in which one particular outcome of a computation becomes known despite the fact that the computer did not run the algorithm.

The controversy arose when Hosten et al.  \cite{Hosten} modified the Jozsa setup claiming to achieve counterfactuality for all outcomes of the computation. In the language of the IFM, Hosten et al. protocol finds both the presence and the absence of an opaque object in a counterfactual manner. The difficulty to define the counterfactuality of the protocol for the case of absence of the object is that we cannot say that the probe was not present because it was not absorbed by the object. Instead, the argument for counterfactuality was that the probe was not present in a particular place because if it were there, it could not have reached the final detector. Vaidman \cite{V07} pointed out that this classical way of considering the location of the quantum probe leads to a contradiction with the symmetry of the quantum description of the probe in the two places, one in which the probe is claimed to be absent and the other in which everyone agrees that it was present. Instead of this classical physics argument, Vaidman suggested an operational definition of the presence of the probe as the place where it left a trace similar to the trace of a probe that was well localized there. According to this definition, the Hosten et al. protocol was not counterfactual.

Salih et al. \cite{Salih13} applied the Hosten et al. idea for ``counterfactual communication'' claiming that in their communication protocol the particle was not present in the transmission channel. Vaidman objected again \cite{V14}, claiming that it is counterfactual only according to the classical physics argument, which cannot be accepted due to associated contradiction, and that it is not counterfactual according to the trace criterion. The controversy continued with numerous publications \cite{V14R,Salih16,Li15,V16C,V16R,V15,Arvid,AGB,AV19,ArBa,Va19,Cao,AhRo,Hance,WCV,LFAZ,SalihNPJ}, but essentially all of them were about counterfactuality in the case of finding that the place is empty, not about the counterfactuality of the original interaction-free measurement of the presence of an object. In particular, when the transmitted bit was 1, corresponding to the blocking of Bob's channel, the trace left in the communication channel was exactly zero, so the protocol was counterfactual according to both definitions. The controversy was only about the case of bit 0, when Bob did not block the channel. In this case, some trace was left in the channel and the discussion was about its size and about the justification to name the protocol ``counterfactual'' when the trace was small but not vanishing.

A separate question in discussions of the protocols, apart from the counterfactuality, was the efficiency of the protocols. Sometimes, the particle did not return to Alice, and these events corresponded to the failure of the protocol. The original IFM protocol had efficiency of only $\frac{1}{4}$, while in the Salih et al. protocol, depending on parameters, efficiency could (theoretically) be arbitrarily close to 1. In the event of failure, the particle was in the transmission channel. It is an essential part of the counterfactual phenomenon, we get information without the particle being in the transmission channel due to the possibility of the particle being there, even though in the legitimate events of the communication protocol, the particle was not there. In the IFM case in the legitimate events the detector in the dark port of the Mach-Zehnder interferometer clicked and in the Salih et al. protocol these were the clicks of detectors $D_1$ and $D_2$ (but not $D_3$).

Griffiths in his paper tried to clarify the controversy by analyzing the presence of the probe in the communication channel, but he missed the target.
Contrary to the literature on this subject, he attributed the term ``counterfactual'' to the issue of  the efficiency of the protocol. He writes: ``The term “counterfactual” in the original SLAZ paper has
the following significance. ... if the number of steps in an SLAZ protocol is
sufficiently large, the magnitude of the amplitude sent through
the channel in each step can be made very small and vanishes
in the limit as the number of steps tends to infinity.''

Griffiths introduced a new criterion that quantified the presence of the probe in the communication channel: ``a well-defined measure of channel usage here called “Cost”, equal to the absolute square of the amplitude sent through the channel''. The problem is that Cost measures the average usage of the communication channel including the cases in which the communication fails and the usage of the channel should not have been taken into account. In the IFM \cite{IFM}, which uses a balanced Mach-Zehnder interferometer, the ``absolute square of the amplitude sent through the channel'' is $\frac{1}{2}$, that is, according to the Cost criterion, the protocol is not counterfactual, in spite of the fact that it {\it defined} the term `counterfactual'. Therefore, Griffith's analysis of the presence of the particle in the transmission channel of the Salih et al. protocol based on Cost might be interesting, but it sheds no light on the question of the counterfactuality of communication protocols.

This work has been supported in part by the U.S.-Israel Binational Science Foundation Grant No. 735/18 and the Israel Science Foundation Grant No. 2064/19.

\bibliographystyle{unsrt}

\vskip 1cm




\begin{thebibliography}{99}


\bibitem{Griff}
R.B. Griffiths,
Multitime quantum communication: Interesting but not counterfactual,
Phys.  Rev. A {\bf 107}, 062219 (2023).



\bibitem{Salih13}
H. Salih, Z.H. Li, M. Al-Amri, and M.S. Zubairy,
Protocol for direct counterfactual quantum communication,
Phys.  Rev. Lett. {\bf 110}, 170502  (2013).



\bibitem{Penrose}
R. Penrose,  {\em Shadows of the Mind}.
 Oxford: Oxford University Press  (1994).

\bibitem{IFM}
A. C. Elitzur,  and L. Vaidman,
 Quantum mechanical interaction-free measurements,
   Found.  Phys. {\bf 23}, 987 (1993).


\bibitem{Jozsa}
R. Jozsa,
Quantum effects in algorithms,
in {\it Lecture Notes in Computer Science},
C. P. Williams, ed. (Springer, London, 1998), Vol. 1509, p. 103.

\bibitem{Hosten}
O. Hosten, M.T.  Rakher, J.T. Barreiro, N.A. Peters, and P.G. Kwiat,
Counterfactual quantum computation through quantum interrogation,
Nature (London) {\bf 439}, 949 (2006).


\bibitem{V07}
L. Vaidman,
 Impossibility of the counterfactual computation for all possible outcomes,
Phys. Rev. Lett. {\bf 98}, 160403 (2007).




\bibitem{V14}
L. Vaidman,
 Comment on ``Protocol for direct counterfactual quantum communication'',
Phys. Rev. Lett. {\bf 112}, 208901 (2014).





\bibitem{V14R}
H. Salih, Z.H. Li, M. Al-Amri, and M.S. Zubairy,
Salih et al. Reply,
Phys. Rev. Lett. {\bf 112}, 208902 (2014).


\bibitem{Salih16}
H. Salih,  Protocol for counterfactually transporting an unknown qubit, Front. Phys.  {\bf 3}, 94 (2016), arXiv:1404.2200 (2014).



\bibitem{Li15}
Z.H. Li, M. Al-Amri, and M.S. Zubairy,
Direct counterfactual transmission of a quantum state,
Phys. Rev. A {\bf 92}, 052315 (2015).





\bibitem{V16C}
L. Vaidman,
Comment on ``Direct counterfactual transmission of a quantum state''
Phys. Rev. A {\bf 93}, 066301 (2016).

\bibitem{V16R}
Z.-H. Li, M. Al-Amri, and M. S. Zubairy,
Reply to ``Comment on ‘Direct counterfactual transmission of a quantum state’”,
Phys. Rev. A {\bf 93}, 066302 (2016).


\bibitem{V15}
L. Vaidman,
Counterfactuality of ‘counterfactual’ communication,
J. Phys. A: Math. Theor. {\bf  48},  465303 (2015).


\bibitem{Arvid}
D. R. M. Arvidsson-Shukur and C. H. W. Barnes
Phys. Rev. A 94,
Quantum counterfactual communication without a weak traces,
Phys. Rev. A {\bf 94}, 062303 (2016).

\bibitem{AGB}
D.R.M. Arvidsson-Shukur, A.N.O. Gottfries, and C.H.W. Barnes,
Evaluation of Counterfactuality in Counterfactual Communication Protocols,
Phys. Rev. A {\bf 96}, 062316 (2017).



\bibitem{AV19}
Y. Aharonov  and L. Vaidman,
  Modification of counterfactual communication protocols that eliminates weak particle traces,
  Phys. Rev. A {\bf 99}, 010103(R)    (2019).

\bibitem{ArBa}
D.R.M. Arvidsson-Shukur and C.H.W. Barnes,
Postselection and counterfactual communication,
Phys. Rev. A {\bf 99}, 060102(R) – (2019).


\bibitem{Va19}
L. Vaidman,
Analysis of counterfactuality of counterfactual communication protocols,
Phys. Rev. A {\bf 99}, 052127 (2019).


\bibitem{Cao}
Z. Cao, Counterfactual universal quantum computation
Phys. Rev. A {\bf 102}, 052413 (2020).

\bibitem{AhRo}
Y. Aharonov  and D. Rohrlich,
What is nonlocal in counterfactual quantum communication?
 Phys. Rev. Lett. {\bf 125}, 260401 (2020).

\bibitem{Hance}
How Quantum is Quantum Counterfactual Communication?
J.R. Hance, J. Ladyman, and J. Rarity,
   Found.  Phys. {\bf 51}, 12 (2021).

\bibitem{WCV}
A. Wander, E. Cohen, and L. Vaidman,
Three approaches for analyzing the counterfactuality of counterfactual protocols,
Phys. Rev. A {\bf 104}, 012610 (2021).


\bibitem{LFAZ}
Z.-H. Li, S.-Y. Feng, M. Al-Amri, and M. S. Zubairy,
Direct counterfactual quantum-communication protocol beyond a single-photon source,
Phys. Rev. A {\bf 106}, 032610 (2022).


\bibitem{SalihNPJ}
H. Salih, W. McCutcheon, J.R. Hance, and J. Rarity,
The laws of physics do not prohibit
counterfactual communication,
npj Quantum Inf. {\bf 8}, 60 (2022).




\end{thebibliography}

\end{document}